\documentclass[journal]{IEEEtran}

\usepackage{algorithm}
\usepackage{algcompatible}
\usepackage{diagbox}
\usepackage{amsmath}
\usepackage{graphicx}
\usepackage{amsmath}
\usepackage{amssymb}
\usepackage{cite}
\usepackage{subcaption}
\usepackage{url}
\usepackage{hyperref}
\usepackage{breakurl}
\ifCLASSINFOpdf
\else
\fi
\AtBeginDocument{%
  \providecommand\BibTeX{{%
    \normalfont B\kern-0.5em{\scshape i\kern-0.25em b}\kern-0.8em\TeX}}}

\begin{document}
%
\title{
Coffee: {\bf Co}st-E{\bf ffe}ctive {\bf E}dge Caching\\ for 360 Degree Live Video Streaming}

%
%

\author{Chen Li, Tingwei Ye, Tongyu Zong, Liyang Sun, Houwei Cao,
        and~Yong Liu,~\IEEEmembership{~Fellow,~IEEE}
\thanks{C. Li, T. Ye, T. Zong, L. Sun and Y. Liu are with the Department
of Electrical and Computer Engineering, New York University, Brooklyn,
NY, 11201 USA e-mail: (\{chen.lee,ty2281,tz1178,ls3817,yongliu\}@nyu.edu).}
\thanks{H. Cao is with New York Institute of Technology, NY, USA e-mail : hcao02@nyit.edu}
}

\maketitle
 
\begin{abstract}
While live 360 degree video streaming delivers immersive viewing experience, it poses significant bandwidth and latency challenges for content delivery networks. Edge servers are expected to play an important role in facilitating live streaming of 360 degree videos. 
In this paper, we propose a novel predictive edge caching algorithm (Coffee) for live 360 degree video that employ collaborative FoV prediction and predictive tile prefetching to reduce bandwidth consumption, streaming cost and improve the streaming quality and robustness. 
Our light-weight caching algorithms exploit the unique tile consumption patterns of live 360 degree video streaming to achieve high tile caching gains. Through extensive experiments driven by real 360 degree video streaming traces, we demonstrate that edge caching algorithms specifically designed for live 360 degree video streaming can achieve high streaming cost savings with small edge cache space consumption. Coffee, guided by viewer FoV predictions, significantly reduces back-haul traffic up to 76\% compared to state-of-the-art edge caching algorithms. Furthermore, we develop a transcoding-aware variant (TransCoffee) and evaluate it using comprehensive experiments, which demonstrate that TransCoffee can achieve 63\% lower cost compared to state-of-the-art transcoding-aware approaches.

\end{abstract}


%
\IEEEpeerreviewmaketitle

\section{Introduction}
A live event can be captured and streamed as a 360 degree video to deliver immersive viewing experience. A viewer dynamically changes her Field-of-View (FoV) by making three-degree-of-freedom (3-DoF) head rotational movements ({\it pitch, yaw, roll}). Meanwhile, live 360 degree video broadcast consumes considerably higher bandwidth than the traditional 2D planar video broadcast.  FoV-adaptive streaming, which partitions a 360 degree video frame into non-overlapping tiles and only streams tiles falling into the viewer's FoV, can  significantly reduce the bandwidth consumption. 
Edge servers are in a unique position to strategically combine communication and computing resources to process and stream 360 degree videos, providing robust and high-quality Quality-of-Experience (QoE) for viewers. Edge-based 360 degree video multicast and caching can support high-throughput and low-latency multi-user delivery. For on-demand 360 degree video streaming, tiles downloaded to an edge server can be cached to serve requests from future viewers. Live 360 degree video streaming can additionally benefit from {\it temporal} locality. Viewers of a live broadcast event are loosely synchronized in time: after a 360 degree video frame is captured, it should be delivered to all the viewers within the maximum tolerable playback latency. Due to heterogeneity of devices and networks, viewers watching the same live event may experience different levels of video quality and incur different playback latencies. To take advantage of the temporal locality of viewers, an edge server can act as a local live streaming proxy to gain the benefits of multicasting. It is usually thought that live content is ``uncacheable". However, the small playback latency gaps among viewers create a niche opportunity to achieve a high caching gain with a low storage cost. Furthermore, by utilizing the computing power on the edge server, an additional caching gain can be achieved by transcoding tiles. For example, a cached tile at one quality level can be transcoded to fulfill requests for the same tile at other quality levels.

In this paper, we investigate predictive edge caching for live 360 degree video streaming, which utilizes \textit{collaborative FoV prediction}, \textit{predictive tile prefetching}, and \textit{transcoding-aware caching} to reduce bandwidth consumption and improve streaming quality and robustness. To the best of our knowledge, this is the first work to design customized edge caching algorithms to reduce live 360 degree video streaming cost. We address two key challenges: (1) \textit{how to design streaming-aware edge caching algorithms to achieve high caching hit ratios with low storage consumption?} and (2) \textit{how to effectively utilize the transcoding capability of edge server to further improve live 360 degree video caching efficiency?} Edge cache boxes have much smaller storage space than the traditional cache boxes, which poses a significant challenge for caching the sheer volume of 360 degree video. To achieve high caching gain, it is critical to accurately predict contents that will be popular in the near future. 
The unique tile consumption patterns of live 360 degree video point to promising directions for  accurate predictive edge tile caching algorithms with low storage cost: (1) Given the video playback start time, one can accurately predict which video frame a viewer will watch at a future time instant; (2) For each 360 degree video frame, a viewer only watches a subset of tiles falling into her FoV, which can be accurately predicted using various FoV prediction algorithms; (3) Viewers of the same live event typically share common viewing interests, which can be leveraged for collaborative FoV prediction; (4) Live video tiles have short life-spans, so there is no need to cache old tiles that all viewers have finished watching; (5) Tile transcoding eliminates the need to store multiple versions of the same tile, as a cached tile can be transcoded to serve requests for the same tile at lower qualities.   

Exploiting these unique opportunities, we make the following contributions towards developing novel edge caching algorithms for live 360 degree streaming:
\begin{enumerate}
\item We predict what a viewer's FoV will be in the near future based on her past FoV trajectory and the FoV trajectories of viewers with a shorter playback latency.

\item With the playback time prediction for a tile, we design a realtime predictive caching score that quantifies how likely a tile will fall into the FoVs of the future viewers.   

\item We further develop transcoding-aware tile caching algorithms that predict the gain of caching a tile at a certain quality level based on not only how likely it will be requested by future viewers, but also how likely it can be transcoded to serve future requests. 


\item Through extensive experiments driven by real 360 degree video streaming traces, we demonstrate that edge caching algorithms specially designed for live 360 degree video streaming can achieve high bandwidth savings with small edge cache space consumption. Our predictive caching algorithms (Coffee and Transcoffee) guided by viewer FoV predictions significantly outperform the state-of-the-art edge caching algorithms.
\end{enumerate}

The rest of the paper is structured as follows. Section~\ref{sec:related} provides an overview of related work. Section~\ref{sec:framework} introduces the edge-assisted live 360 degree video streaming framework. Section~\ref{sec:predictive} develops predictive edge caching algorithms guided by collaborative FoV prediction. Section~\ref{sec:transcodingcosteffective} explains how we design the transcoding-aware edge caching algorithm. Section~\ref{sec:evaluation} presents the evaluation results. Finally, Section~\ref{sec:conclusion} outlines future work.

\section{related work}
\label{sec:related}
Edge content caching helps deliver contents at low-latency and high-throughput by caching popular contents, so that the traffic in the core/back-haul network can be significantly reduced. However, edge cache boxes have smaller storage space than the traditional CDN servers, and serve just a small number of users, whose aggregate content request patterns are much more dynamic. Naive caching strategies such as Least-Recently-Used (LRU) or Least-Frequently-Used (LFU) are insufficient to get good caching performance. To deal with this challenge, many existing studies proposed various algorithms including both heuristic-based~\cite{wei2021wireless,abolhassani2021single} and learning-based methods~\cite{song2020learning,fan2021pa,wang2020intelligent,ye2021joint,li2023predictive} to achieve good edge-caching performance. In~\cite{tian2022dima,zong2022cocktail}, deep reinforcement learning is also used for edge caching. 
Previously, some studies focused on edge caching-assisted on-demand video streaming~\cite{han2019proactive}. 
However, there are only very few studies on edge caching for live 360 degree video multicasting~\cite{sun2020flocking,pantelis2021tilebased}. In~\cite{pantelis2021tilebased}, LSTM network is used to predict the evolution of the popularity of video tiles in future GOPs in tile-based live 360 degree video streaming but no transcoding is considered when making the caching decision. 
Live 360 degree video multicast can benefit from the caching and transcoding capabilities of edge servers. Trans-coding is usually applied in 360 streaming systems. ~\cite{yang2022collaborative} leverage a DRL-based model to make caching and transcoding decisions to reduce delay, transmission cost and quality mismatch level with limited cache size and computational power.
In~\cite{xiao2022transcoding}, multi-agent RL is proposed for a transcoding-enable edge caching model to minimize communication and transcoding latency with a constrained transcoding power. 
However, the cost of transcoding can not be ignored. In our work, TransCoffee minimizes total cost of communication and transcoding in live 360 degree video streaming system using a lightweight model.
In addition, FoV prediction is a necessary part for 360 degree video streaming, no matter for on-demand or live systems. Impressive research have been accomplished for FoV prediction~\cite{li2023spherical,yaqoob2021combined}, and edge-assisted delivery~\cite{Nahrstedt_sarkar2021l3bou,zhong2022multi,wang2023colive} of 360$^o$  video recently. Collaborative FoV prediction is proposed in~\cite{ban2018cub360} which predict a user's FoV by other users' viewing directions, but they are designed for on-demand 360 degree video streaming. Collaborative FoV prediction for live 360 degree video streaming was investigated in~\cite{sun2020flocking} under the framework of ``user flocking". However, it was not used to design special edge caching algorithms and FoV prediction accuracy will drop for long-term prediction.
The focus of our work is to design novel edge caching algorithms exploiting long-term collaborative FoV prediction and edge-based tile transcoding to achieve high bandwidth savings in live 360 degree video streaming with low edge cache storage consumption. To the best of our knowledge, this is the first work to use edge-caching to minimize total communication and transcoding cost in live 360 degree video streaming.



\begin{figure}[htbp]
\centerline{\includegraphics[scale=0.4]{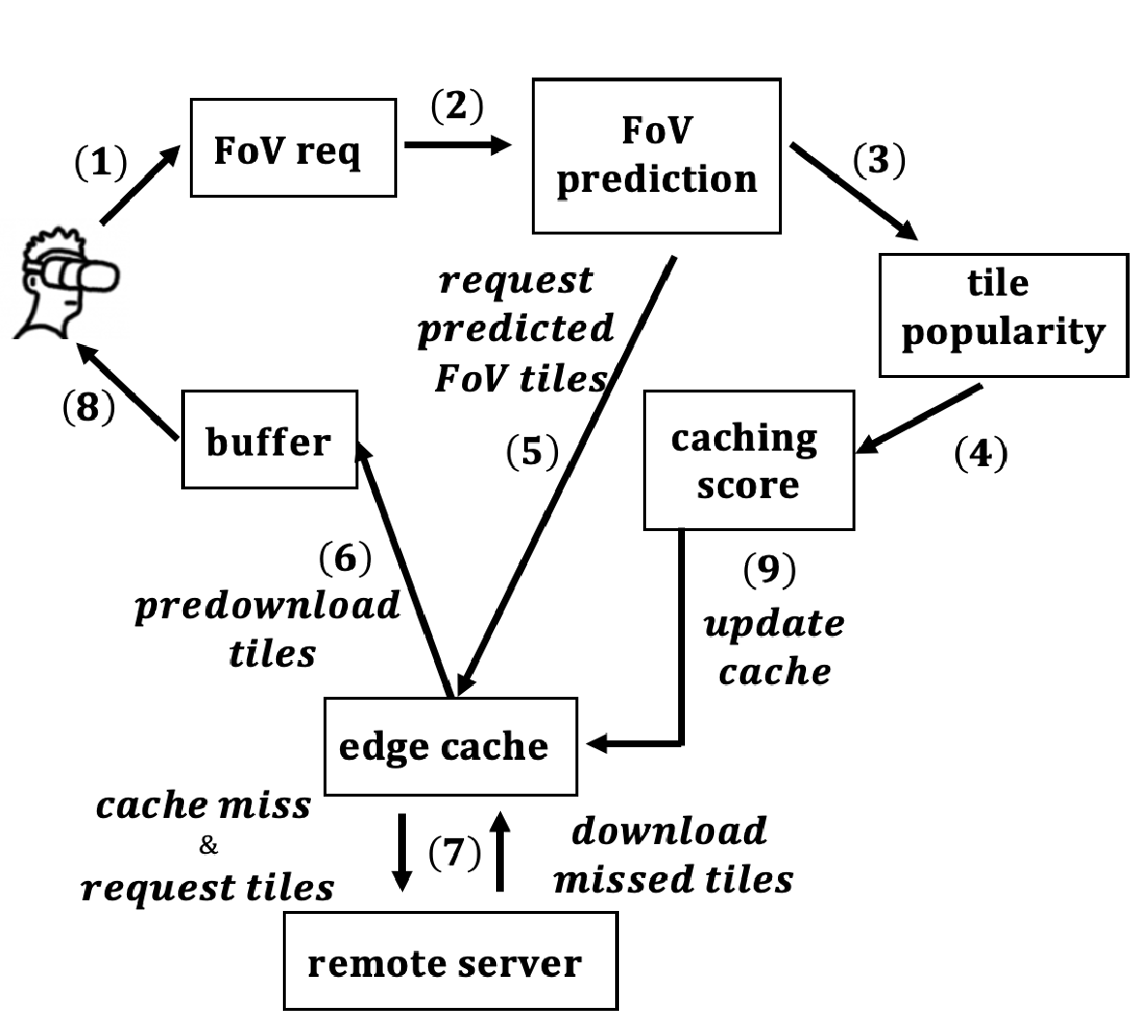}}
\vspace{-3mm}
\caption{System Workflow of Edge-assisted Live 360 Degree Video Streaming.}
\label{fig:system}
\end{figure}

\section{Edge-Assisted Live 360 Degree \\
Video Streaming}
\label{sec:framework}
Before diving into the detailed caching algorithm designs, we present an overview of how live 360 degree video streaming can benefit from caching and transcoding on edge boxes. A live event is captured as a 360 degree video whose frames are temporally grouped into Group-of-Pictures (GoPs). Each GoP is spatially partitioned into tiles, each of which contains video along a certain direction. Each tile is encoded into multiple quality levels. Adaptive streaming dynamically selects the quality level to be streamed to match the available bandwidth. We focus on a group $\mathcal V$ of viewers of the live event served by a common local edge box. When viewer $i$ joins the live broadcast, she chooses an initial playback latency $l_i$ and streaming buffer length $b_i < l_i$ based on her network condition. In general, viewers with stronger/more stable network connections can afford to select shorter $l_i$ and $b_i$. At time $t$, viewer $i$ plays tiles captured at time $t-l_i$, while downloading tiles captured at time $t-l_i+b_i$. We define $d_i \triangleq l_i-b_i$ as the tile download lag of viewer $i$. To reduce bandwidth consumption, only tiles inside the viewer's FoV are downloaded. To facilitate this, viewer $i$'s FoV for video at $t-l_i+b_i$ is predicted based on her FoV trajectory for the video up to $t-l_i$, and the tiles in the predicted FoV will be downloaded. Instead of requesting tiles directly from the original streaming server, viewer $i$ uses the edge box as her streaming proxy. If a requested tile is already in the edge cache, the tile can be directly streamed from the edge box, without generating traffic in the core/backbone network. If the requested quality version is lower than the cached version, transcoding will be conducted and the transcoded tile will be streamed to the viewer. Fig.~\ref{fig:system} illustrates how different streaming and caching functions work together: 
\begin{enumerate}
    \item A viewer joins the live 360 degree video broadcast with initial playback latency and streaming buffer length. 
    \item The viewer's FoV trajectory is fed to a FoV prediction module to predict her future FoV.
    \item Based on the FoV prediction, the future tile popularity is estimated for the viewer.
    \item Tile caching scores are calculated according to the tile popularity, as detailed in Section~\ref{caching score} and transcoding-aware variant in Secton~\ref{sec:transcodingcosteffective}. 
    \item Tiles in the predicted FoV will be requested. Tile downloading requests are directed to the edge server, which will check whether the requested tile is in cache or not. 
    \item If the requested tile is cached (potentially after transcoding), the tiles will be sent to the viewer's local streaming buffer.
    \item If the requested tile is not cached, a new request is sent to the remote server and download the missed tile from remove server to local edge cache. 
    \item The viewer plays tiles from the head of the buffer. We assume that each viewer will always download a base layer tile with a low resolution, but it will encompass the entire 360 degrees scene.
    \item The updating module will consult the caching scores generated by the caching policy to update the cache. It notes that whenever a new tile request is received, the associated caching scores will be updated.
\end{enumerate}

\section{Edge Tile Caching based on Collaborative FoV Prediction}
\label{sec:predictive}

\subsection{Live 360 Degree Video Tile Caching}
Among a group $\mathcal V$ of viewers, the video tile captured at time $t$ will be requested first by the viewer with the shortest tile download lag at time $t+min_{i \in \mathcal V} d_i$. The tile will be streamed from the remote streaming server and can be cached in the local edge server to serve the other viewers. The last request for this tile will be generated by a viewer at time $t+max_{i \in \mathcal V} d_i$. A naive tile caching solution is to store the tile for the duration of $max_{i \in \mathcal V} d_i-min_{i \in \mathcal V} d_i$ so that all subsequent requests for the tile can be served from the edge cache. We call this base caching algorithm as \texttt{Caching-All}. We will compare its performance with our proposed caching algorithms in Fig.~\ref{fig:no transcoding video0} in Section~\ref{sec:evaluation}. A major disadvantage of \texttt{Caching-All} is that, since different viewers have different interests, if a tile is not viewed by any subsequent viewers, the cache space is wasted. To efficiently use edge cache space, we predict the probability of a newly downloaded tile being watched by future viewers, and only cache tiles that are predicted to be highly popular. We will use collaborative viewer FoV predictions to generate accurate estimates of future tile popularity.

\subsection{Collaborative Viewer FoV Prediction} 
\begin{figure}[htbp]
\centerline{\includegraphics[width=0.8\linewidth]{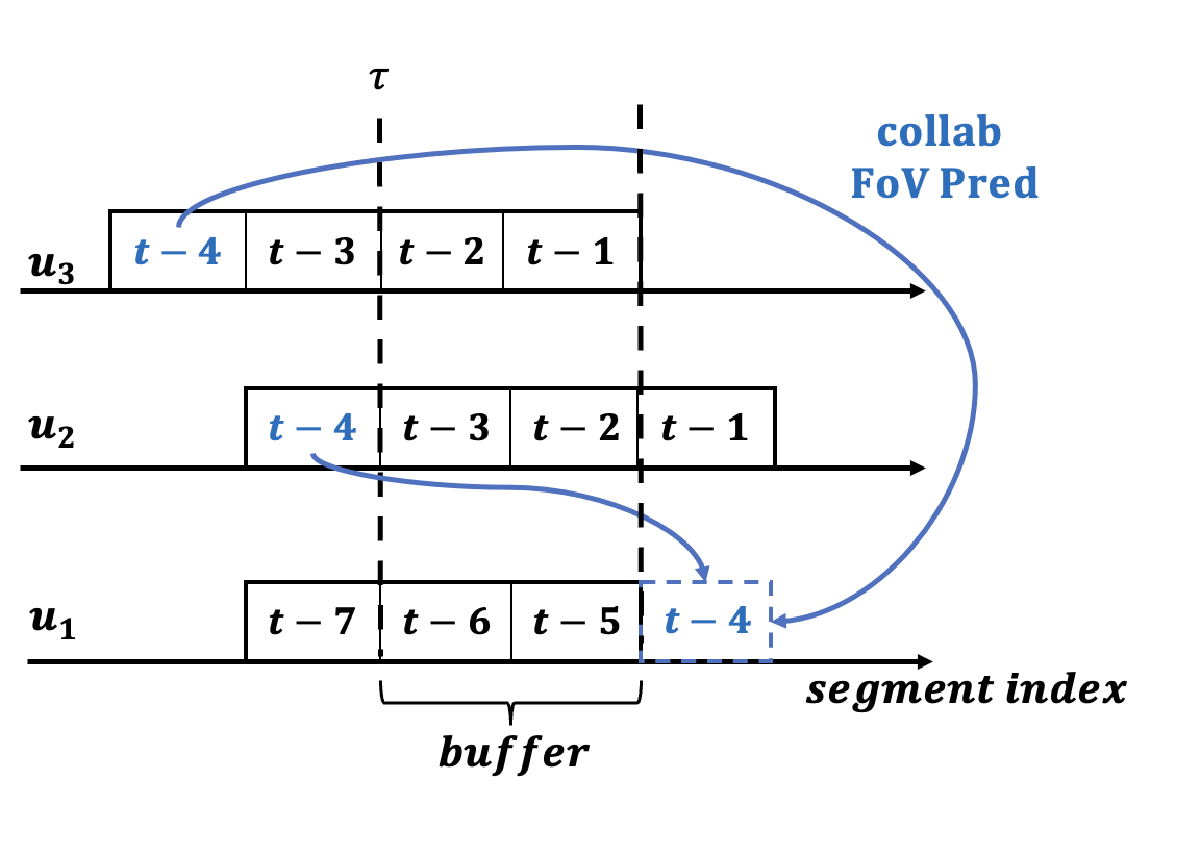}}
\caption{In a streaming system, if the current time is $\tau$, $u_3$ is the user with the shortest playback latency and will watch segment $t-2$. $u_2$ will watch the segment $t-3$, and $u_1$ will watch the segment $t-6$. If the buffer length is 2, $u_1$ needs to predict FoV for the segment $t-4$, which has been watched by $u_2$ and $u_3$. Therefore, we can use the FoVs of $u_2$ and $u_3$ on the segment $t-4$ for collaborative FoV prediction. The weight used to combine the FoV information from $u_2$ and $u_3$ is based on the similarity between $u_1$ and these two users.}
\label{fig:streaming buffer effect}
\end{figure}

During the live streaming session, the viewer's FoV is recorded for each frame. We quantify the viewer's interest for each tile of a GoP as the average overlap ratio of the tile with the viewer's FoV for each frame in the GoP. At a given time $t$, we have the ground-truth of the viewer $i$'s interests for video up to $t-l_i$. To pre-fetch video at $t-d_i$, we need to predict the viewer's interest for video at $t-d_i$, with the streaming buffer length $b_i=l_i-d_i$ being the prediction interval. In addition to the viewer $i$'s own past tile interests, we also have the ground-truth viewing interests for viewers who have already played video segment $t-d_i$, namely, all viewers with playback latency $l_j<d_i$. Since viewers of the same event often share common viewing interests, those viewers' tile interests are also used to predict the target viewer's tile interests. As illustrated in Fig~\ref{fig:streaming buffer effect}, we adopt a collaborative FoV prediction algorithm from \cite{sun2020flocking}. The prediction has two parts: 1) Truncated Linear Prediction (TLP) based on the target viewer's past FoV trajectory for video up to $t-l_i$, 2) a weighted sum of similar viewers' tile interests for video segment $t-d_i$. The similarity between a pair of viewers is measured by the distance between their FoV trajectories for the last one second using the Dynamic Time Warping (DTW) algorithm.

\subsection{Tile Request Time Prediction}
\label{subsec:request time}
Due to the limited size of the edge caching memory, it is essential to place the correct content at the appropriate time. We not only forecast whether a tile will be requested by a viewer, but also when the viewer will request it, so that we can adjust the cached content in a timely manner. Predicting the time of content requests is usually a difficult task for general edge content caching. In live streaming, a viewer usually downloads and plays video segments sequentially. Given a download delay $d_i$, viewer $i$ will download tiles from the video segment $t-d_i$ at time $t$ with a stable network connection. We are concentrating on a cost-efficient edge caching system, similar to~\cite{maniotis2021tile,papa2019tile,maniotistile2020}, where we assume that users have a reliable network connection and all requests can be fulfilled from either the edge caching server or the remote server.  By combining the prediction of viewer $i$'s tile interests for video segment $t-d_i$ with this, we can generate high-quality real-time caching scores for all active tiles. 

\subsection{Tile Caching Score based on FoV Prediction}
\label{caching score}
\begin{figure}[htbp]
\centerline{\includegraphics[width=0.75\linewidth]{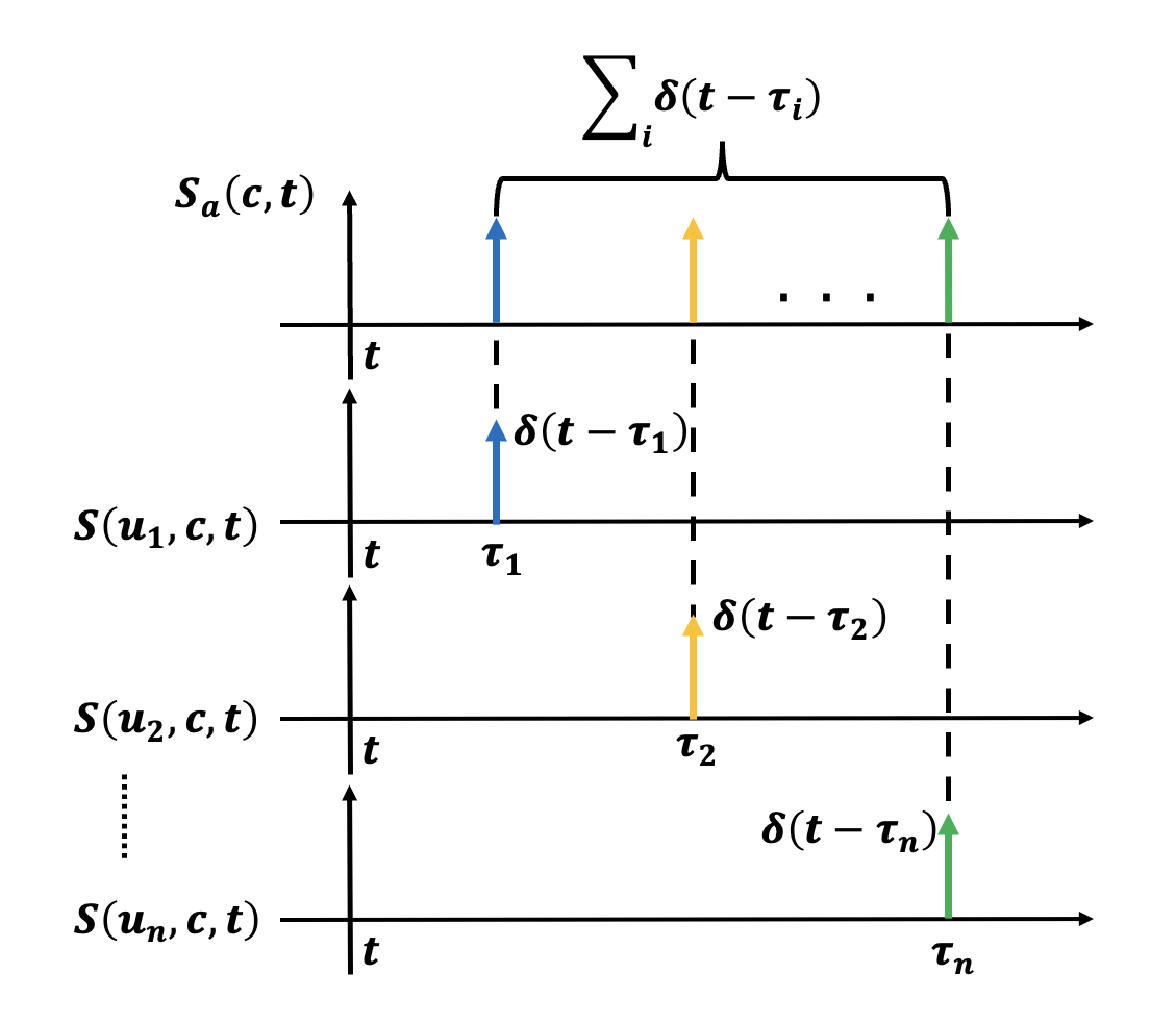}}
\caption{The aggregation of scores from $n$ viewers for a tile $c$ yields a caching score $S_a(c,t)$ at the current time $t$. Each viewer has a single contribution $S(u,c,t)$ to the score.}
\label{fig:aggregation}
\end{figure}

For the group of viewers who are using a same edge server, we predict the popularity of a tile within a video segment by aggregating the predicted viewing interests of all viewers who have not watched this video segment. If we predict that a viewer will watch tile $c$ at future time $\tau$, we will generate a per-user predictive caching score as:  
\begin{equation}
\label{predictivescore2}
    \mathcal S(u,c , t)= \delta(t- \tau),
\end{equation}
where $\delta(\cdot)$ is the unit impulse function as illustrated in Fig.~\ref{fig:aggregation}.
To get the caching score for a tile, we aggregate the predictive caching scores from all viewers who have not watched this segment, which is $\mathcal{V'}$:
\begin{equation} 
\label{aggregationscore}
 \mathcal S_a(c, t) = \sum_{u \in \mathcal V'}   S(u,c,t)
\end{equation} 

\begin{figure}[htbp]
\centerline{\includegraphics[width=0.7\linewidth]{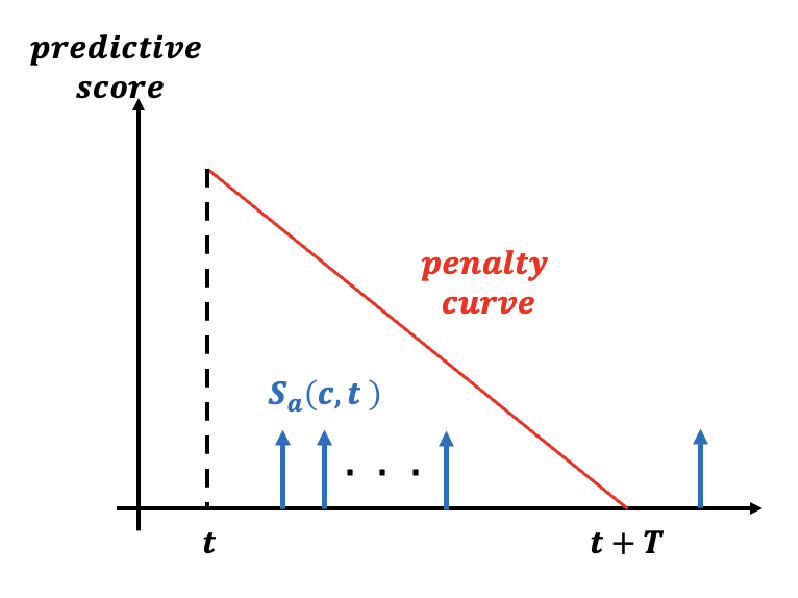}}
\caption{Once an aggregated predictive caching score has been obtained, a penalty curve will be applied to penalize scores at extended prediction horizons, as accuracy may decrease at these points.}
\label{fig:penalty}
\end{figure}

Considering that the accuracy of FoV forecasting diminishes as the prediction horizon increases, we compute the ultimate caching score as the integral of $S_a(c, t)$ weighted by a time-decaying penalty function:
\begin{equation} 
\label{eq:tile_popularity}
 \mathcal S_f(c) = \int_{t}^{t+T} S_a(c,\tau) P(\tau)  \,d\tau,  
\end{equation} 
 where a linearly decaying function $P(\tau) = T - \tau + t$ is used in our experiments. 
An illumination is shown in Figure.~\ref{fig:penalty}. Note that $S(u,c , t)$ will be updated after each new tile request from each viewer. So the final caching score $S_f(c)$ is also dynamically updated as new FoV predictions are generated. Another potential advantage of the penalty function is that we do not rely on the predicted FoV beyond the time horizon $T$, because predictions beyond $T$ have no impact on $S_f(c)$.

\subsection{Predictive Tile Caching}
Based on the predictive tile caching score $S_f(c)$, we design a caching policy for live 360 degree video streaming, as demonstrated in Algorithm.~\ref{alg:constant-size caching}.  
\begin{algorithm}
{\fontsize{9pt}{9pt}\selectfont
\caption{Predictive Edge Tile Caching}
\label{alg:constant-size caching}
  \begin{algorithmic}[1]
 	\STATEx  {\textbf{Input:} user $u$ requests content $c_{i}$, i indicates item/tile, time is $t$}
	\STATEx {\textbf{Output:} refreshed items in cache.}
        \IF { tile $c_{i}$ in cache}
		\STATE {$hit++$}                            
        \ELSE
            \STATE{add $c_{i}$ in cache}
        \ENDIF
        \STATE{remove all tiles of segments $\tau$ in cache, where $\tau < t - d_{max}$ }
        
        \WHILE{total tile size in cache $>$ Cache Size}
            \STATE{get the lowest $S_f(c_{j})$ from cache}
            \STATE{remove $c_{j}$ from cache}
        \ENDWHILE  
\end{algorithmic}}
\end{algorithm}
The predictive caching score $S_f(c)$ is employed in cache replacement. When the cache is full, tiles with the lowest $S_f(c)$ will be evicted. Additionally, all tiles of segment $t-d_{max}$ will be taken out of the cache at time $t$, since no viewer will ask for it in the future. In reality, different viewers have different access networks with varying bandwidths and rendering devices with different maximum resolutions. Different viewers may request the same tile at different quality levels. Without transcoding on the edge, the tile at each quality level is considered as a distinct item in caching score update and cache replacement using Algorithm~\ref{alg:constant-size caching}.

\section{Transcoding-aware Caching}
\label{sec:transcodingcosteffective}
Recently, edge servers have been increasingly utilized for content delivery. They are equipped with computing resources to support real-time transcoding for 360 degree videos. Despite the growing computational power of edge servers, the cost of transcoding can still not be ignored. For example, by AWS Elemental MediaConvert Pricing~\cite{AWStranscodingprice}, the per minute transcoding cost for 4K video can be \$0.045. 
On the other hand, the storage cost is relatively low compared to the bandwidth and transcoding costs~\cite{AWScacheprice}. For instance, a cache node cache.m5.12xlarge, with 48 vCPUs and 157.12 GB of memory, costs \$3.744 per hour. However, a single high-resolution 360 degree video can reach 3.29 Gbit/s ~\cite{Huawei_Report}. At the same time, extra large caching memory is slow and edge servers usually have limited storage capacity. Therefore, it is essential to consider how to do edge caching with a limited cache size, especially as file sizes become larger in the future. 

We present TransCoffee, a transcoding-aware, cost-effective edge caching policy for live 360-degree video streaming that takes into account transcoding costs. The TransCoffee caching score $S_{Tcf}$ is transcoding-aware, meaning that the caching gain for each tile at one quality level depends on whether the same tile has been cached at other quality levels. For instance, if a high quality tile is cached, caching lower versions of the tile may not be necessary since they can be obatined through transcoding from the cached high quality tile. As a result, we need to estimate the additional gain of caching a tile at certain quality level given the cached quality levels. We resort to the Shapley Value \cite{shapleyvalue} to fairly divide the profits from cooperation when multiple versions of a tile are required to be cached. This is an average of the marginal contributions from all possible permutations of cached versions. However, in a real-time caching system, we need an instant gain of each tile, so we use the instant marginal contribution~\cite{marginalcontributionnet} from the Shapley value. If the cache system is not real-time, the expectation of marginal contribution may be necessary.

First, we have $M$ quality versions for each tile, the size of version $r$ is $w(r)$ and let's say: $w(0)<w(1)<...<w(r)<...w(M)$. For each tile $c$ at resolution $r$, we use Equation~\ref{equ:transcoding cost score} to represent the final transcoding-aware caching gain for $c_r$, which evaluates the unit caching gain of caching this tile $c_r$ at time $t$. It is noted that all the variables related to caching gain are time dependent; however, for the sake of simplicity, the time variable $t$ is omitted in the notation, for example, $S_{Tcf}(c_r,t)$ is written as $S_{Tcf}(c_r)$.


\begin{equation}
\label{equ:transcoding cost score}
S_{Tcf}(c_r) = \frac{G_{Tcf}(R_{in} \cup c_r)- G_{Tcf}(R_{in})}{w(r)},
\end{equation}
where $G_{Tcf}(X_c)$ represents how much gain we can get if we cache the tile set $X_c$, where $X_c$ is an arbitrary set of resolution levels for a tile $c$. $R_{in}^{(c,r)}$ is the set of the other cached resolution levels except level $i$. We mark it as $R_{in}$ for simplification. Similarly, $R_{out}^{(c,r)}$ is the set of the other non-cached resolution levels except level $r$, and we mark it as $R_{out}$. For one tile $c$, $R_{in} \cup R_{out} \cup c_r = \{c_0, c_1, ..., c_M\}$. \\
The cost of transcoding, as demonstrated in the test in \cite{xiao2022transcodingcost} and the pricing of commercial transcoding services such as AWS Elemental MediaConvert~\cite{AWStranscodingprice}, is largely dependent on the target resolution level. Therefore, we define the cost of transcoding to target level $k$ as $T_k$ regardless of the original resolution. Additionally, the bandwidth overhead is another factor to consider and we define the unit downloading cost as $B_c$, which indicates the amount of money spent on downloading one unit size of the content. \textbf{The total caching gain after adding the resolution level $r$ to the currently cached resolution levels} is:

\begin{equation}
\label{equ:Gtcf union higher non empty}
\begin{split}
G_{Tcf}(R_{in} \cup c_r) &= \sum_{j \in R_{in}} P_j * w(j) * B_c \\
                         &+ P_r * w(r) * B_c \\
                         &+ \sum_{k\in R_{trans}} P_k * (w(k) * B_c - T_k),  
\end{split}
\end{equation}
where $P_r$ is the popularity of tile $c$ at resolution level $r$ and it can be predicted by $\mathcal S_f(c)$ in Equation.~\ref{eq:tile_popularity}. The first two terms are the bandwidth cost saving from serving directly from cached tiles without transcoding. The last term is the net cost saving resulted from serving requests after transcoding cached tiles, which is the bandwidth cost saving minus the transcoding cost. $R_{trans}$ is the set of resolution levels that are currently not cached, but can be served by transcoding the cached resolution levels:
\begin{equation}
R_{trans} \triangleq \{ c_m: c_m \in R_{out}, m < pivot \} 
\end{equation}
where $pivot = \max$ resolution level in $\{R_{in} \cup c_r \}$. We call $R_{trans}$ the set of resolutions covered by transcoding.  

Similarly, \textbf{the caching gain of the other cached tiles except level $r$} is: \\
\begin{equation}
\label{equ:Gtcf rin higher non empty}
    \begin{split}
    G_{Tcf}(R_{in}) &= \sum_{j \in R_{in}} P_j * w(j) * B_c \\
                         &+ \sum_{k\in R_{trans'}} P_k * (w(k) * B_c - T_k) 
    \end{split}
\end{equation}
where 
\begin{equation}
R_{trans'} \triangleq \{ c_m: c_m \in R_{out} \cup c_r, m < pivot' \} 
\end{equation}
and $pivot' = \max$ resolution level in $\{R_{in}\}$.\\
Compared to Equation~\ref{equ:Gtcf union higher non empty}, version $i$ cannot be served directly from the cache, but if a higher version is in the cache, version $r$ can still be served after transcoding. Overall, the set of transcoding-covered resolutions may shrink if we remove the current resolution $c_r$. Therefore, the caching gain of adding level $r$ to the current cache is:
\begin{equation}
\label{caching gain of tile ci transcoding cost score }
    \begin{split}
S_{Tcf}(c_r) * w(r) &= P_r * w(r) * B_c  \\
            &+ \sum_{k \in \{ R_{trans} \} } P_k * ( w(k) * B_c - T_k ) \\
            &- \sum_{k \in \{ R_{trans'} \}} P_k * ( w(k) * B_c - T_k )
    \end{split}
\end{equation}

To further simplify $S_{Tcf}(r_r)$, we define $R^{r+}$ as the set of cached tiles with a resolution level higher than $r$, and $R^{r-}$ as the set of cached tiles with a resolution level lower than $r$. Thus, $R_{in}$ is the union of $R^{r-}$ and $R^{r+}$.\\
\textbf{If $R^{r+} \neq \emptyset$, }which means that there is a higher level of tile in the cache and then:
\begin{equation}
    R_{trans'} = R_{trans} \cup c_r 
\end{equation}
\begin{equation}
\label{final transcoding cost score higher is not empty}
    \begin{split}
S_{Tcf}(c_r) &= P_r * B_c - P_r * ( B_c - \frac{T_r}{w(r)})\\
            &= P_r * \frac{T_r}{w(r)}
    \end{split}
\end{equation}
\textbf{If $R^{r+} = \emptyset$,} which means $c_r$ is the highest level, then
\begin{equation}
      R_{trans} - R_{trans'} \triangleq \{ c_m: c_m \in R_{out}, pivot' < m < r \}
      \end{equation}

\begin{equation}
\label{transcoding cost score}
    \begin{split}
S_{Tcf}(c_r) &= P_r * B_c  \\
            &+ \frac{1}{w(r)} \sum_{k \in \{ R_{trans} - R_{trans'} \} } P_k * ( w(k) * B_c - T_k )
    \end{split}
\end{equation}

Besides, if we use real transcoding and download cost, sometimes, transcoding cost can larger than downloading cost, so we only execute transcoding when $ B_c > \frac{T_r}{w(r)}$. So the final unit transcoding gain is: 

We have:
\begin{equation}
\label{final transcoding cost score higher is empty}
    \begin{split}
S_{Tcf}(c_r) &= P_r * B_c - P_r * \max \{( B_c - \frac{T_r}{w(r)}),0 \}\\
    \end{split}
\end{equation}
So
\begin{equation}
\label{transcoding cost score consider TD ratio}
    \begin{split}
&S_{Tcf}(c_r) = P_r * B_c  \\
    &+ \frac{1}{w(r)} \sum_{k \in \{ R_{trans} - R_{trans'} \} }  \max \{ P_k * ( w(k) * B_c - T_k ), 0 \}
    \end{split}
\end{equation}

As long as we possess a TransCoffee score $S_{Tcf}(c_r)$, we will update the cache by utilizing Algorithm~\ref{alg:Multiple Version Tile caching policy}. Note that to simplify the expression, we omit subscript $t$ in TransCoffee score $S_{Tcf}(c_r)$, which is dynamic and will be updated over time for real time edge caching. Besides, the latency of transcoding is usually insignificant and can be accommodated in the user's streaming buffer\cite{xiao2022transcodingcost}, such as two seconds.

\begin{algorithm}
{\fontsize{9pt}{9pt}\selectfont
\caption{Transcoding-aware Edge Caching Update Policy}
\label{alg:Multiple Version Tile caching policy}
  \begin{algorithmic}[1]
 	\STATEx  {\textbf{Input:} user $u$ requests content $c_{i,r}$, i indicates item/tile, where r indicates resolution of this tile. $T_{c}$ and $D_{c}$ indicates transcoding and downloading cost of tile $c$. Real-time caching score $S(c,t)$ at time $t$ of tile $c$.}
	\STATEx {\textbf{Output:} refreshed items in cache.}
        \IF{$c_{i,r}$ $\in$ cache}
            \STATE{$hit++$}
        \ELSIF {tile $c_{i,h}$ $\in$ cache, where $h > r$ and $T_{c_{i,r}}<D_{c_{i,r}}$}
            \STATE{transcode $c_{i,h}$ to $c_{i,r}$}
            \STATE{add $c_{i,r}$ in cache}
            \STATE {$hit++$}            
        \ELSE
            \STATE{download $c_{i,r}$ from remote server and add it to cache}
        \ENDIF
        \WHILE{total tile size in cache $>$ Cache Size}        
            \STATE{get the $c_{j,k}$ with lowest $S_{Tcf}(*)$ from cache}
            \STATE{remove $c_{j,k}$ from cache}
        \ENDWHILE     
\end{algorithmic}}
\end{algorithm}

\section{Performance Evaluation}
\label{sec:evaluation}
We simulate edge caching assisted live 360 degree video streaming to assess the effectiveness of the proposed algorithms.
\subsection{Dataset}
We utilize a public 360-degree video dataset from~\cite{wu2017dataset}, which consists of 48 viewers (24 males and 24 females) watching sphere videos. The first three videos from different categories are chosen and their details are presented in Table~\ref{tab:allvideo}. Unless otherwise specified, single-video experiments are conducted in video0. We assume that each video is divided into GOPs of 1 second and each GOP is spatially partitioned into 30 tiles (5*6). The viewer's request for each GOP is generated based on her average interest ratio for the 30 frames in the GOP. A viewer will request a tile if it has overlap with the predicted FoV. We assign a playback latency of $l_i$ to all users, with a uniform distribution over a period of 20 seconds. Following the 360-degree video resolution in~\cite{tileresolution}, we set the entire 360-degree video coded in 100, 500, 1000, 1500, 2000, and 2500 Mbps.

\begin{table}[htbp]
    \centering
    \begin{tabular}{|c|c|c|}
        \hline
        {\bf Video Index} & {\bf Video Name} & {\bf Video Category} \\
        \hline
        video0 & Conan360-Sandwich &Performance\\
        \hline
        video1 & Freestyle Skiing &Sports \\
        \hline
        video2 & Google Spotlight-HELP &Film\\
        \hline
    \end{tabular}
    \caption{We use three videos from different categories, where viewers have different FoV patterns.}
    \label{tab:allvideo}
\end{table}
\subsection{Live Streaming System Configurations}

In live streaming, tiles have short life spans, meaning that after a certain period of time, no viewers will request them. The playback latency for streaming services, such as those offered by sports networks and platforms like ESPN, NBC Sports, and CBS Sports, can be as long as 40 seconds~\cite{sportslatency}, while for cable and satellite, it can be around 5 seconds. In our experiments, we set the maximum viewer download time lag as $d_{max}=20$ second. A tile is considered inactive after it has been viewed for the first time and 20 seconds have elapsed. We set the cache size from 0 to a sufficiently large size, at which the cache memory can store all the active tiles. We normalize the cache size by the memory needed to cache all active tiles at the  highest resolution: given the longest download time lag is $d_{max}$, the number of tiles in each GOP is $M$, and the highest resolution tile size is $S$, the maximum cache size is $F = d_{max} * M * S$. In our experiments, $F=187.5GB$ for one video. Unless otherwise specified, the cache size in a figure represents the ratio over the maximum cache size. The buffer length of the live streaming system is 2 seconds, unless otherwise noted. The $T$ in Equation.~\ref{eq:tile_popularity} is set as 15 seconds + buffer length.


\subsection{FoV Prediction Accuracy}
\label{sec:prediction accuracy}
In this section, we evaluate the accuracy of FoV prediction with different prediction approaches. We evaluate these three approaches:
\begin{itemize}
    \item Truncated linear prediction(TLP) is a simple linear prediction method that predicts a viewer's future FoV trajectory by extrapolating from the recent monotone linear pattern in the nearest truncated FoV history of the viewer.     
    \item ColPB: Collaborative FoV prediction with boundary constraint, a FoV prediction method that predicts a user's next tile based on the collaborative FoV prediction algorithm of~\cite{sun2020flocking}. FoV of the target viewer for an upcoming video segment is predicted using the target viewer's own trajectory as well as the trajectories of viewers who have watched this video segment. The prediction gives a higher weight to the target viewer's own trajectory with 
    \begin{equation}
    \label{eq:self-weight}
    \max \left\{\frac{1}{1 + sum(similarities)}, 0.8\right\},
    \end{equation} 
    where $sum(similarities)$ is the sum of similarity of all viewers who have watched the segment. The weight given to the other viewers is one minus the weight to the target viewer.  
    \item ColP-Long: We customize ColPB~\cite{sun2020flocking} for long-term FoV prediction necessary for 360-degree live video caching. We remove the lower bound of 0.8 in (\ref{eq:self-weight}).  This is because a longer prediction horizon is needed than that of ColPB. TLP-based self-prediction does not perform well when predicting at a long horizon. By removing the lower bound of 0.8, if the summation of similarities with the front viewers is high, more reliance is placed on the front viewer's information rather than the viewer's self-trajectory for collaborative prediction. Additionally, instead of prefetching the top k tiles as in ColP, we select a FoV whose coverage of tiles can maximize the total tile weight within that FoV to calibrate the tile distribution prediction.
\end{itemize}
\begin{figure}[htbp]
\centering{
\centerline{\includegraphics[scale=0.21]{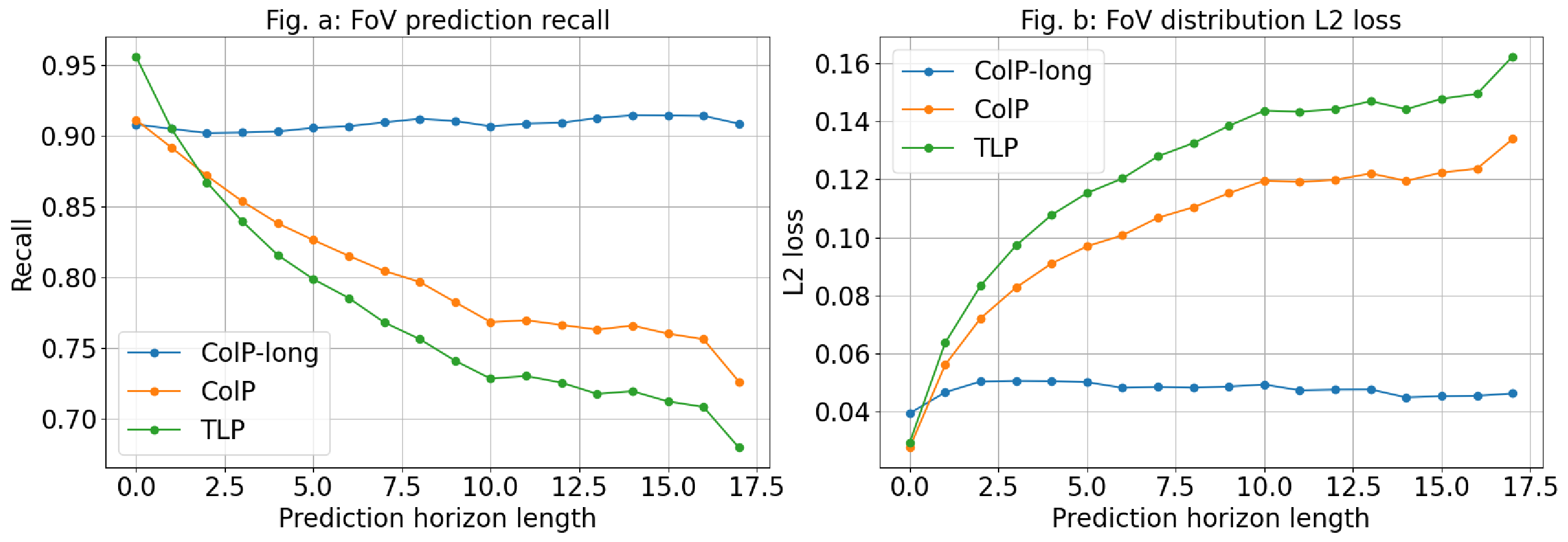}}
\caption{The y-axis in Fig.a is the ratio of overlap between the actual FoV of the viewer and the predicted FoV at different prediction horizon lengths. In Fig.b, we calculate the L2 loss between the predicted tile probability and the tile distribution in the viewer's actual FoV. If a tile is partially overlapped with the FoV, the value for the tile is the fraction of tile pixels in the FoV. If a tile is completely in the FoV, the value is 1. After normalization, we obtain a distribution of all tiles for the user's actual FoV. We compare the L2 loss between these two distributions in Fig.b. Since we have a buffer length of 2 seconds, we only consider the prediction accuracy after the buffer length.} 
\label{fig:predictionaccuracy_video0}}
\end{figure}


 We conducted experiments to compare the accuracy of three FoV prediction methods. We used the target viewer's FoV trajectory in the last GoP to predict the future trajectory for TLP. Figure~\ref{fig:predictionaccuracy_video0} shows the evaluation of the overlap ratio with the viewer's actual FoV in video 0. The x axis is the length of the prediction horizon. Figure~\ref{fig:predictionaccuracy_video0} (a) illustrates the fraction of tiles in the viewer's actual FoV that can be covered by the downloaded tiles based on the FoV predictions.
 We can see ColP-Long has consistently higher overlapping ratios than TLP and ColPB at all prediction horizons. This is important to generate accurate predictive caching scores. 


\subsection{Caching without Transcoding}
\label{sec:notranscoding}
We begin by utilizing our prediction module and pre-user caching score for non-transcoding scenarios. We set different resolutions for tiles to compare with the transcoding cases in the next section. Each viewer has a specific resolution on her device and can only request that resolution. The edge server does not have transcoding ability and considers different versions of the same tile as distinct content items. Algorithm~\ref{alg:constant-size caching} is employed for non-transcoding caching. We compare our algorithm with three live 360 degree video caching baselines using different tile caching scores. 

\begin{itemize}
    \item LRU-live: a caching replacement policy that evicts content based on the time elapsed since the last request for a tile and with similar tile eviction after $d_{max}$. It eliminates a tile from the cache after the longest download wait time $d_{max}$ (20 seconds in this case). As after 20s, the tile will never be asked for\footnote{We also tested on LFU-live, which yielded much poorer results than LRU-live, as the lifespan of live video tiles is brief.}.
    \item NOC-live, as described in~\cite{zhou2021caching}, is designed to minimize the difference between the performance of an online learning algorithm and the best dynamic policy that could have been chosen in hindsight. To adapt to fast popularity changes of live video tiles, the step size delta is set to 0.9. After $d_{max}$ seconds, tiles will no longer be requested and will be removed from the live streaming cache.
    \item Latency-FoV (LF)*. In \cite{sun2020flocking}, a 360 degree live video streaming caching policy was proposed, which is based on the Least Recently Used (LRU) approach. It groups users based on their FoV trajectory similarity, and then marks users with the longest latency witin each group, and ignores the content downloaded by those users. In contrast, we put all users in the same group, and mark users with the longest 25\% latency and not cache content downloaded by these users, since only a few future users can benefit from them.
    \item Coffee: This is our caching scheme without transcoding. It eliminates content based on the dedicated caching score, which is determined by the collaborative FoV prediction approach. Specifically, ColP-long is chosen for FoV prediction as it is more effective than TLP and ColP. The caching score is $ \mathcal S_f(c)$ in Equation~\ref{eq:tile_popularity}.
\end{itemize}

\begin{figure}
    \centering
    \includegraphics[scale=0.3]{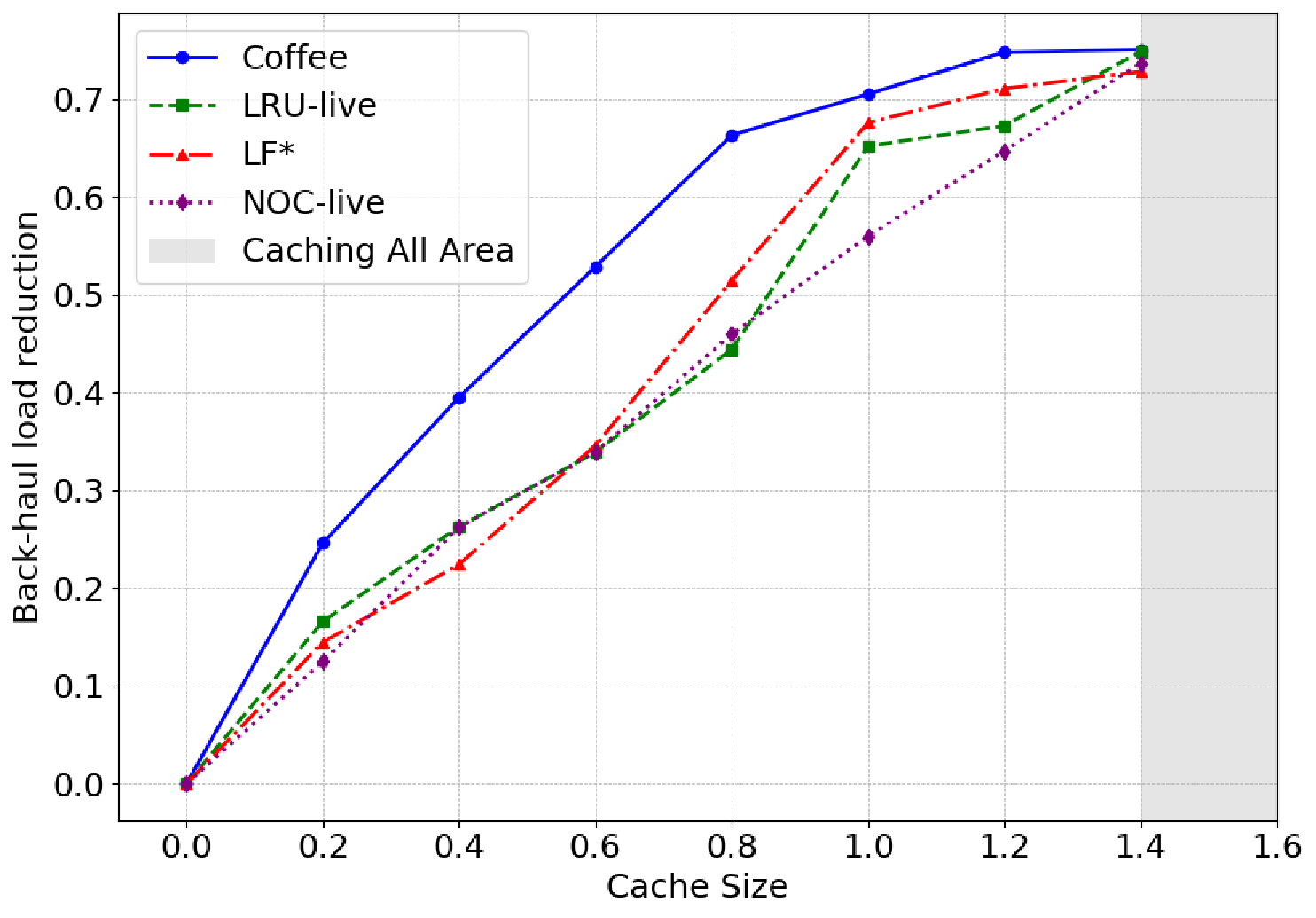}
    \caption{The back-haul load reduction ratio for Video0 without trans-coding is compared for different cache sizes. The x axis of the graph ranges from 0 to 262GB, with the unit cache size being 187.5GB. Our method outperforms all benchmarks, and can almost double the back-haul load reduction, due to the precise predictive caching scores generated by the per-viewer FoV prediction. For example, Coffee will reduces the back-haul traffic up to 76\% compared to LF* when cache size is 0.4. The area on the right hand side indicate the Caching-All, which means cache size is large enough to cache all tiles.}
    \label{fig:no transcoding video0}
\end{figure}

\begin{table*}[htb]%
\centering
\caption{Back-haul load reduction for Ours, NOC-live, LRU-live on different cache size and videos.}
  \label{table:no transcoding all videos}
{
\begin{tabular}{||c||c|c|c||c|c|c||c|c|c||}
\hline
\diagbox{}{Scenarios}  & \multicolumn{3}{c||}{\bfseries Cache Size=0.4} & \multicolumn{3}{c||}{\bfseries Cache Size=0.8} & \multicolumn{3}{c||}{\bfseries Cache Size=1.2} \\
  \hline
  \cline{2-10}
Method &Video0 &Video1 &Video2 &Video0 &Video1 &Video2 &Video0 &Video1 &Video2 \\
\hline
LRU-live  &0.2628 &0.2443 &0.2447 &0.4441 &0.4521 &0.4165 &0.6729 &0.6346 &0.6565 \\
NOC-live  &0.2622 &0.2547 &0.2372 &0.4596 &0.4292 &0.4157 &0.6472 &0.5979 &0.5864 \\
LF*       &0.2243 &0.2183 &0.2047 &0.5145 &0.4839 &0.5238 &0.7107 &0.6932 &0.6762 \\
Coffee    &{\bf 0.3951} &{\bf 0.4228} &{\bf 0.3879} &{\bf 0.6632} &{\bf 0.6397} &{\bf 0.6428} &{\bf 0.7483} &{\bf 0.7024} &{\bf 0.7289} \\
\hline
\end{tabular}
}
\end{table*}

The reduction ratio in back-haul bandwidth, which is the bandwidth consumption in the connection between the edge server and the original content server, with different cache sizes for video0 is depicted in Fig.~\ref{fig:no transcoding video0}. Our method consistently outperforms the baselines. When the cache size is unlimited, all caching policies will have the same performance by caching everything they encounter. Nevertheless, since the high resolution of 360 videos can be very large, it is still important to reduce the back-haul bandwidth using as little cache space as possible. Our approach is aware of which resolution a viewer will request, and this also boosts its performance since different versions of a tile are distinct content items without transcoding. We have similar results in other videos, as summarized in Table.\ref{table:no transcoding all videos}. 

Furthermore, it only takes $0.019067$ second to finish each prediction of user content interest (10 seconds horizon) and the cache score aggregation. Our predictive algorithms are efficient and can be implemented in real time.


\subsection{Transcoding-aware Caching}
\label{sec:transcoding}
We evaluate the effectiveness of our Transcoffee system with various caching policies by utilizing transcoding-aware Algorithm~\ref{alg:Multiple Version Tile caching policy}.

Baseline:

\begin{itemize}
    \item TransCoffee: We propose a Transcoding-aware cost-effective edge caching technique that eliminates content based on the anticipated popularity of the tile. This approach is derived from the ColP-long FoV prediction approach and a specialized transcoding-aware tile score.
    \item Coffee: The Coffee method does not have the capacity to transcode. It applies the same FoV prediction model and the caching score without the transcoding term, as indicated in Equation \ref{eq:tile_popularity}.
    \item LRU-live-T: It is an LRU-live-based transcoding-variant method and caching policy based on Algorithm~\ref{alg:Multiple Version Tile caching policy}. When the cache is full, contents are evicted using their LRU scores.
    \item E-Transcoding\cite{sun2020flocking}: Always cache the highest-level tile and run transcoding all the time. This approach requires only one level of tile in the cache, and we use ColP-long to predict the future popularity of tiles, based on which we use (\ref{eq:tile_popularity}) to calculate their caching scores for eviction.
    \item LF*-T. We use a transcoding version of modified Latency-FoV in \cite{sun2020flocking}, and cache  updating follows 
 Algorithm~\ref{alg:Multiple Version Tile caching policy}.
    \item NOC-live-T: It's a NOC-live-based caching method~\cite{zhou2021caching} with transcoding resources, and follows the cache updating Algorithm~\ref{alg:Multiple Version Tile caching policy}.
    \item ETE\_0C: End-to-end baseline with zero cache size, like \cite{mahzari2018fov}, this approach just delivers tiles to users directly without cache to bound the caching performance.
\end{itemize}

\subsubsection{Transcoding and downloading cost}
We assess the performance of TransCoffee by considering the transcoding and downloading costs from the latest AWS Elemental MediaConvert Pricing~\cite{AWStranscodingprice} for transcoding and Amazon EC2 On-Demand Pricing~\cite{AWSbandwidthprice} for downloading cost. The cost of transcoding (per minute) for each tile depends on the resolution: 4K is \$0.045, HD is \$0.0225 and SD is \$0.0113. The downloading cost is based on the coded tile size and is \$0.09 per GB. Additionally, from \cite{xiao2022transcodingcost}~\cite{AWStranscodingprice}, transcoding cost is mostly determined by the target resolution level. To fully evaluate the effect of the transcoding cost, we also have different transcoding/downloading cost ratios(T/D ratios), which will be discussed in Section~\ref{subsection:TD ratio}.

\subsubsection{Streaming cost}
\begin{figure*}
    \centering
    \includegraphics[scale=0.55]{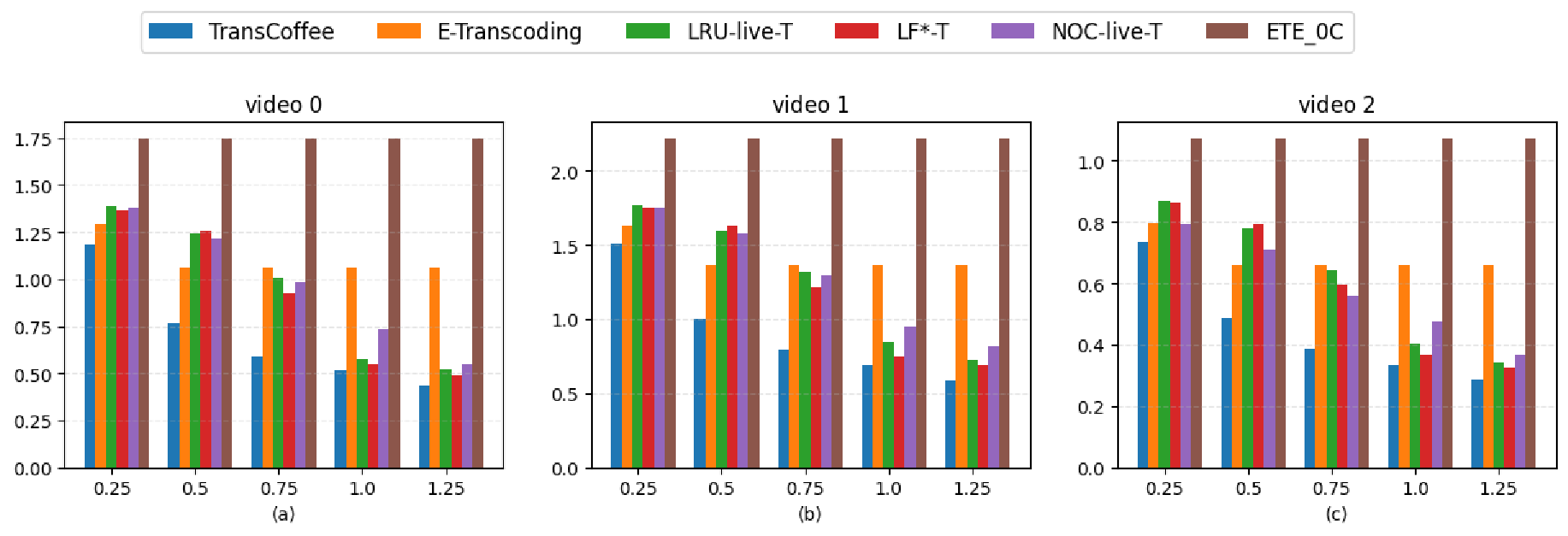}
    \caption{The total cost per user for each method is shown in the subfigures, with the x-axis representing the cache size and the y-axis representing the cost in dollars. The cost of transcoding and downloading is based on the prices of AWS. TransCoffee constantly outperforms all state-of-the-art baselines on different cache sizes and different videos, and the cost reduction compared to LF*-T can reach 63\% when the cache size is 0.5 on video 0.}
    \label{fig:allvideso_t}
\end{figure*}



\begin{figure*}[h]
    \centering
    \begin{subfigure}[b]{0.33\textwidth} 
        \includegraphics[width=\textwidth]{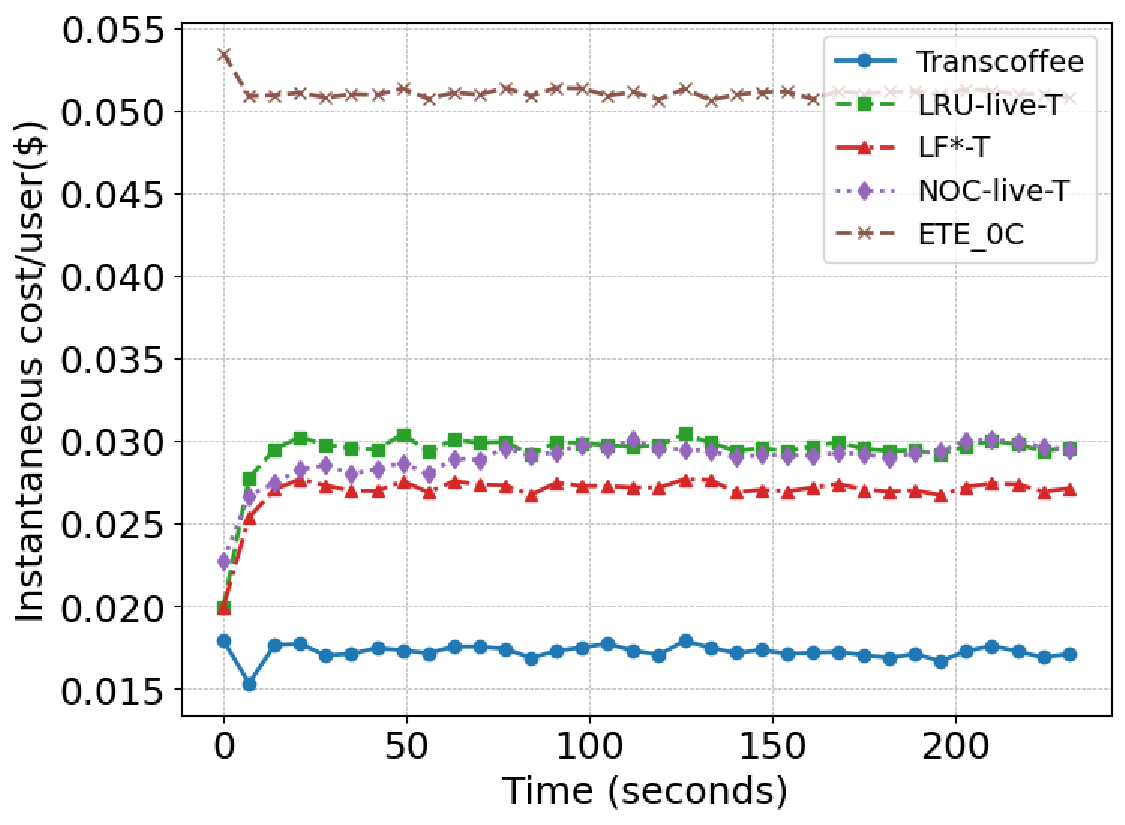}
        \caption{}
        \label{fig:cost with time aws}
    \end{subfigure}
    \hfill
    \begin{subfigure}[b]{0.31\textwidth}
        \includegraphics[width=\textwidth]{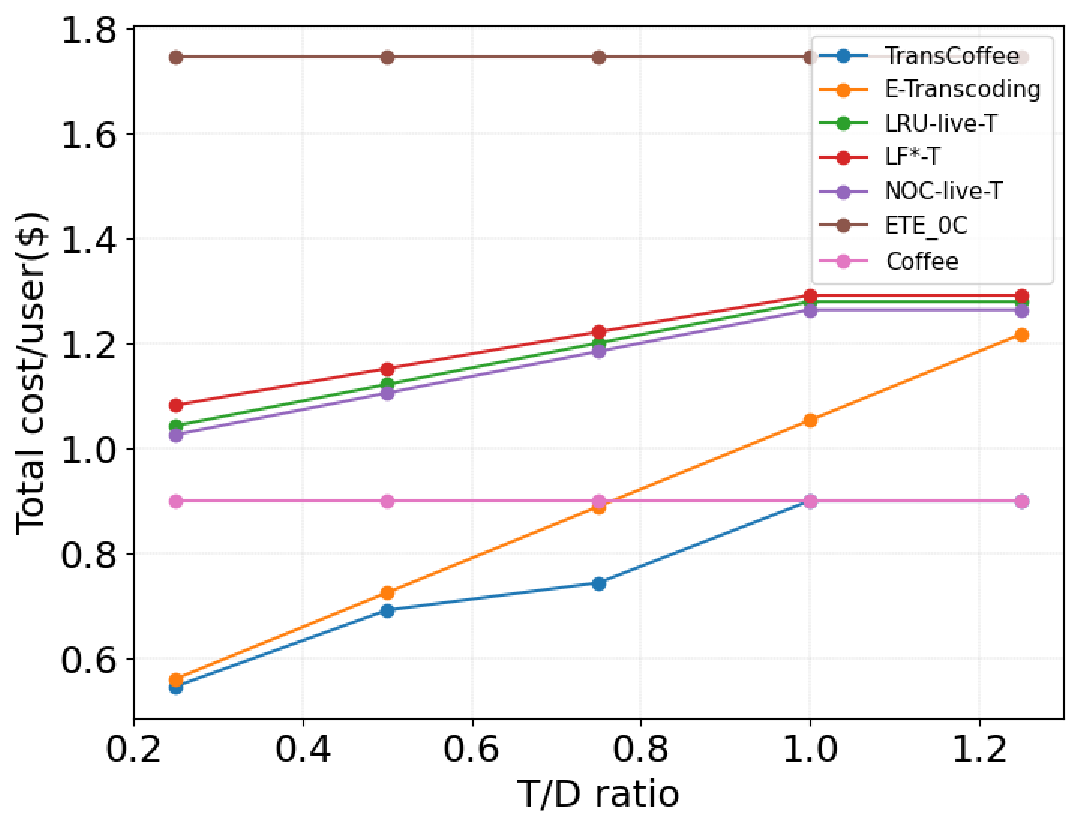}
        \caption{}
        \label{fig:T_D_R_aws}
    \end{subfigure}
    \hfill
    \begin{subfigure}[b]{0.30\textwidth}
        \includegraphics[width=\textwidth]{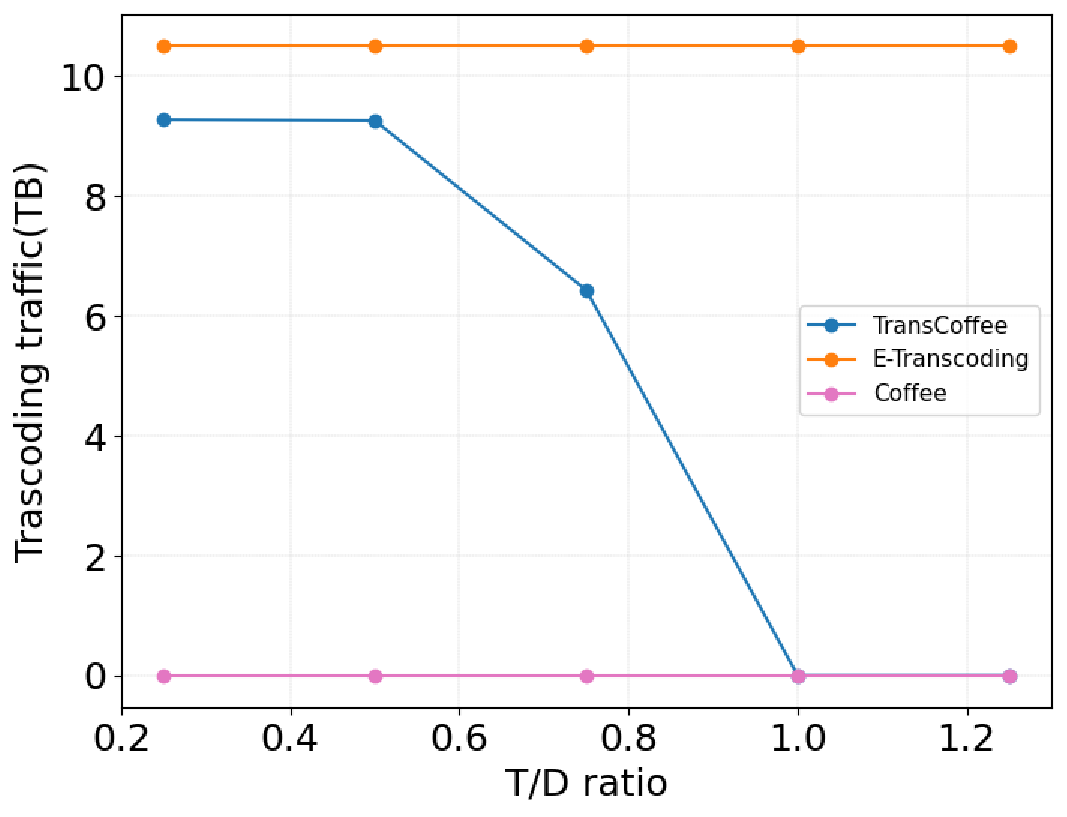}
        \caption{}
        \label{fig: transcoding traffic T_D_R_aws}
    \end{subfigure}
    \caption{Fig. (a) shows the total cost over time for video 0 when using AWS pricing and a cache size of 0.75. Fig. (b) displays the total cost per user for different ratios of transcoding and downloading costs. Fig. (c) illustrates the transcoding traffic with the transcoding and downloading cost ratio.}
    \label{fig:main}
\end{figure*}

\begin{figure}[htp]
    \centering
    \includegraphics[scale=0.4]{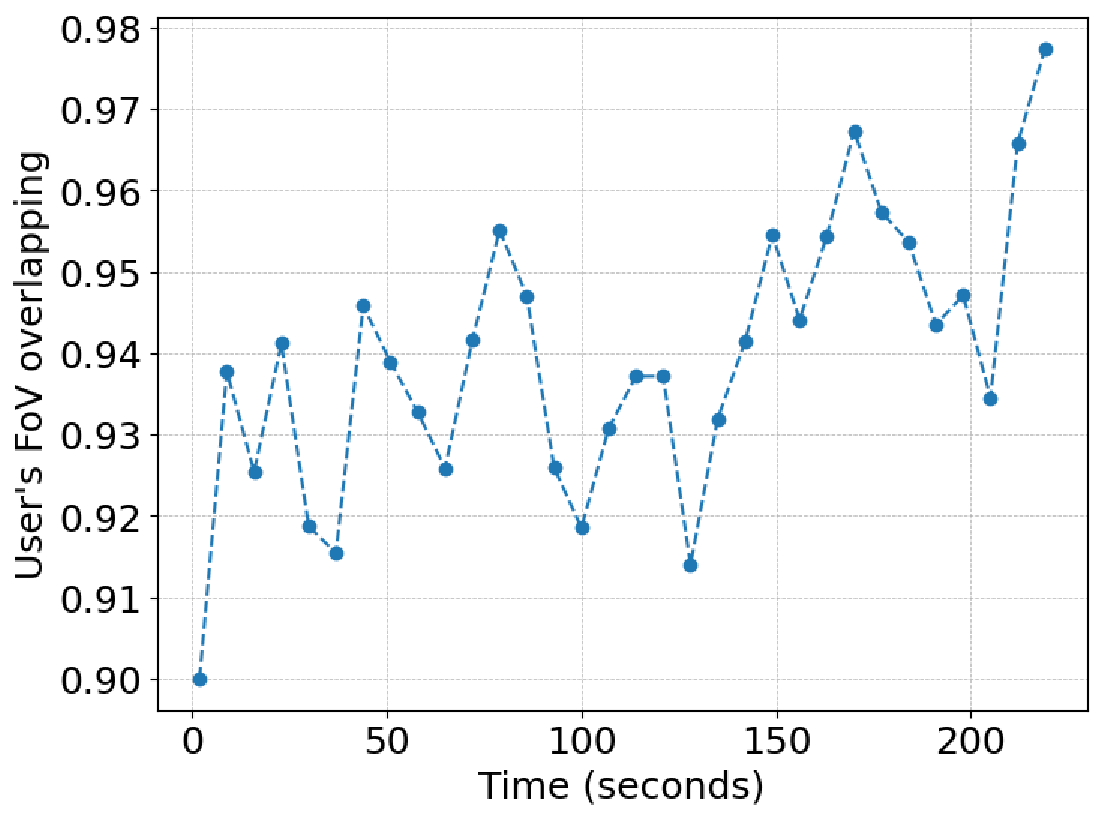}
    \caption{The viewer's interest hit byte ratio over time for video 0, with the x-axis representing the stream time in seconds and the y-axis representing the fraction of tiles in the viewer's actual FoV that can be rendered using the tiles downloaded in the streaming buffer.}
    \label{fig:interest_hit_ratio}
\end{figure}

We evaluate caching performance with varying cache sizes for different videos in Fig.~\ref{fig:allvideso_t}. TransCoffee outperforms the other benchmarks. Fig.\ref{fig:cost with time aws} illustrates the cost with time for all methods when the cache size is 0.5 and the streaming cost is based on the AWS price. TransCoffee manages to achieve the lowest cost throughout the streaming process.

\subsubsection{Performance on different T/D ratio}
\label{subsection:TD ratio}



We evaluated the effect of transcoding cost and the robustness of our method by conducting extra experiments on different transcoding and downloading cost ratios. We fixed the downloading cost using the AWS bandwidth price and proportionally changed the transcoding cost to get different T/D ratios. In Fig.~\ref{fig:T_D_R_aws}, it is shown that the total cost per user with different T/D ratios when the cache size is 0.5. TransCoffee has the lowest streaming cost compared to all the baselines, and its performance can converge towards E-Transcoding when the transcoding cost is low (always transcoding) and to Coffee (no transcoding) when the transcoding cost is higher than the downloading cost. Between 0 to 1, TransCoffee has a much lower cost compared to the other methods, demonstrating the robustness of our method with different T/D cost ratios. In Fig.~\ref{fig: transcoding traffic T_D_R_aws}, we evaluated transcoding traffic with different T/D ratios when the cache size is 0.5. Transcoding traffic of TransCoffee decreases with increasing T/D ratio. When the transcoding cost is 0, TransCoffee utilizes transcoding as much as possible to save streaming cost, and when the transcoding cost is equal to the downloading cost, no more transcoding will be used to save streaming cost. The adaptation of transcoding traffic is consistent with Fig.~\ref{fig:T_D_R_aws}, indicating that TransCoffee can adaptively utilize transcoding resources by referring to our transcoding-aware caching score, thus reducing the total streaming cost.


\subsection{Hit Ratios for Different Groups of Viewers}
The prediction module of our system predicts the popularity of tiles based on viewers with shorter playback latency. This allows the subsequent viewers to benefit from the pre-downloaded tiles of the front viewers. Table~\ref{tab:usergrouphitratio} shows the hit-byte ratio among different viewer groups. The viewer's index is based on the ranking of their playback latency in the live video broadcast. User0 is the one with the shortest playback latency, and user47 has the longest playback latency. The hit-byte ratio for the first group (user 0-5) is low because they are the ``front-runners" in the live broadcast and there are not many pre-downloaded tiles in the cache to serve them. The following user groups (e.g. user 6-17) can have a high hit-byte ratio if the correct tiles are cached. However, for the later user groups, their hit-byte ratio can be low again. This is because they are at the end of the live video broadcast, and the tiles they request will not be requested in the future, and the 'caching gain' for these tiles will be low. To make the best use of cache space, our caching policy chooses to sacrifice these tiles to reserve space for the tiles that will become `more popular' in the near future. The smaller the cache size, the more obvious this phenomenon is. This makes sense because our goal is to reduce the whole back-haul traffic by using caching for all of the users together, and requests with a lower hit-byte ratio can still be satisfied by sending requests to the remote server.
\begin{table}[h]
    \centering
    \begin{tabular}{|c|c|c|c|}
        \hline
        \diagbox{user group}{cache size}  & 0.1 & 0.3 &0.6 \\
        \hline
        0-5 &  0.2485 &0.3214  &0.3339 \\
        \hline
        6-11 &  0.7794  &0.9201 &0.9496 \\
        \hline
        12-17 &  0.67638 &0.9537 & 0.97807 \\
        \hline
        18-23 & 0.0248 &0.9500  &0.9949 \\
        \hline
        24-29 & 0 &0.8772  &0.9854 \\
        \hline
        30-35 & 0 &0.7041  & 0.9749\\
        \hline
        36-41 & 0 &0.0551  &0.9856 \\
        \hline
        42-47 & 0 &0  &0.3232\\
        \hline
    \end{tabular}
    \caption{Tile hit ratios for users with different playback latencies at different cache sizes.}
    \label{tab:usergrouphitratio}
\end{table}

\subsection{Viewer's Interest Hit Byte Ratio}
We assess the viewer's interest hit byte ratio for a portion of video0. Figure~\ref{fig:interest_hit_ratio} illustrates the fraction of tiles within a viewer's actual FoV that can be rendered using tiles downloaded in the streaming buffer. The average viewer's interest hit byte ratio is greater than 93\%, indicating that we can provide a satisfactory viewing experience to viewers using pre-fetched tiles in our experiments.

\label{subsection:Flocking Caching variable-size}



\section{Conclusion}
\label{sec:conclusion}
This paper has demonstrated that the predictability of a viewer's FoV and the sequential tile download patterns of live 360 degree video streaming can be exploited to design accurate predictive edge tile caching algorithms that achieve high bandwidth savings with low storage consumption. We further showed that transcoding-aware caching algorithms can improve edge caching efficiency. Experiments conducted with real 360 video streaming traces of 48 viewers demonstrated that our algorithms can reduce the bandwidth consumption of live 360 degree video streaming by up to 76\% compared to streaming without caching. Higher bandwidth reductions are expected with more viewers served by the same edge cache box. Additionally, our transcoding-aware caching algorithms(TransCoffee) outperformed the state-of-the-art edge caching algorithm by up to 63\%. 


\bibliographystyle{IEEEtran}
\bibliography{tmmref,software}

\end{document}